# On the Performance of P2P Network: An Assortment Method

**Yuqing Zhou**

Department of Mathematics

Wuhan University of Technology, Wuhan, China

Email: yuqing.zhou@live.whut.edu.cn

**Abstract**

Peer to Peer (P2P) systems have grown dramatically in recent years. The most popular type of P2P systems are file sharing networks, which are used to share various types of content over the Internet. Due to the increase in popularity of P2P systems, the network performance of these systems has become a very important issue in the design and realization of these networks. Hence, the performance of the P2P has been improved. This paper will suggest the following methods for the improvement of the P2P systems: Method-1: Improve the P2P routing by using a sandwich technique. Method-2: Improve the search performance by introducing a new search based on the super peer. Method-3: Improving the search by introducing a ranking algorithm based on the knowledge database. The system demonstrates that the methods introduced here have the improved efficiency compared to the previous methodologies. So, the results show that the performance of the P2P systems have been improved by using the above methods, hence the traffic can be reduced.

**1. Introduction**

Peer-to-Peer (P2P) system has become one of the hottest research topics, its excellent characteristics of fully decentralized control and self-organizing make it attractive for some particular applications. Routing algorithm has great influence in P2P applications. Current routing algorithm concentrates on creating well organized network architecture to improve the routing performance. However, for each routing procedure, system returns location information of the requested file only and the characteristics of system workload are considered. Routing performance is critical for P2P networks to achieve high performance [1].

Peer-to-Peer systems are one of the effective ways of communicating and sharing data equally. The new age protocols are able to route efficiently with faster download, better stability and a good user interface. However, determining the destination peer in the first place is not always trivial. Selection of right peer for the given query is difficult. Especially the distributed knowledge management setting suffers from the behavior of selfish peers which are autonomous and require some incentive for their service.

P2P has become a popular application among the Internet. The searching of the content and file sharing is perhaps the most popular application. An application uses flooding as their prevailing resource location method for content searching and file sharing. The basic problem with the flooding mechanism is that it creates a large volume of unnecessary traffic in the network mainly because a peer may receive the same queries multiple times through different paths.

This paper has been organized as follows: 1. the performance of the routing in the P2P networks has been improved by using the sandwich technique. This can be explained in the method-1. 2. In the method-2, matching of user query should be enhanced and most exact result to be obtained by the user at a single search by using the ontology based search. 3. The performance of the searching is improved using super peer based search with improved searching technique (A novel approach) is explained in the method-3.

## 2. Proposed Methods and Results

### Method1: (Sandwich Technique)

Using sandwich method, the HDHTR and SDHTR are combined based on the following criteria: 1. creating a P2P ring where routing tasks are first executed in lower level ring before they go to higher level thus reducing routing overhead. 2. A super node based routing algorithm which reduce the average routing latency and cost [2, 3]. Hence it reduces the routing overhead and average routing latency, which thereby reduce the traffic.

The nodes are first hierarchically arranged by the default algorithm. The nodes within the ring will communicate each other without the help of super node. If two nodes from different rings are communicate only through the super node. Using the sandwich method, we combine the HDHTR with SDHTR. Using the simulation, we proved that the sandwich method improve the performance of the peer to peer network which indirectly reduces the traffic. The Figure-1 shows the nodes in different rings with their super node. Nodes available within a ring will communicate each other directly. If nodes are in different rings must communicate each other using their super node only.

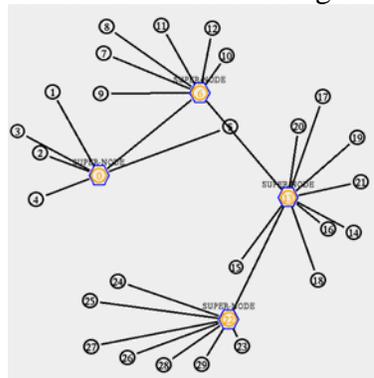

Figure – 1. Different rings (group of nodes) with their super node

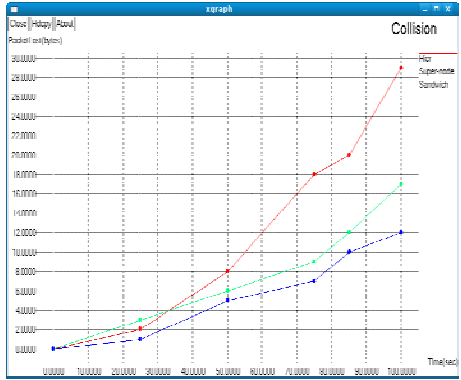

Figure – 2. The Comparison of Collision (based on packet lost in time) occurred

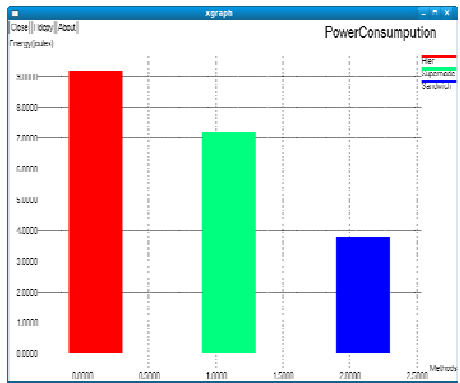

Figure – 3. The Comparison of Power consumption (based on joule)

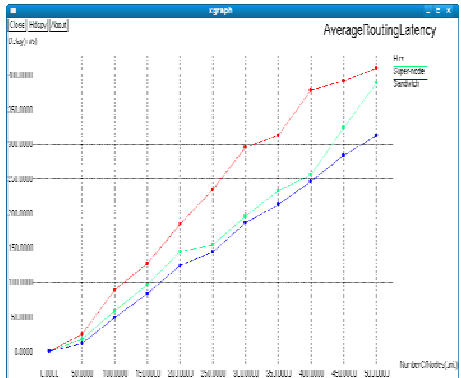

Figure – 4. The Comparison of Average Routing Latency

In the figure-2, the collision occurred when the methods are implemented is compared based on the packet lost in seconds are showed. In the figure-3, powers consumed by the methods are compared based on the joules. In the figure-4, the average routing latency of the methods are compared and showed.

SDHTR does not separate system nodes as clients and servers as in Client Server Architecture, nor does it treat all system nodes equally as other P2P systems. It classifies peers into super nodes and ordinary nodes according to their different resources and capabilities. HDHTR, each P2P ring, the members have the equal responsibilities for workloads in this ring. A routing procedure is executed in lower layer P2P rings before it goes up, and eventually reaches the global P2P ring. By taking a large portion of routing hops in lower layer P2P rings which have smaller link latency between any two nodes inside these rings. Using the sandwich methodology, the method use a super node based routing algorithm and P2P ring to reduce the routing overhead and average routing latency, which thereby reduce the traffic.

**Method2: (Knowledge based Peer Ranking)**

In proposed data model, the ontology based selection method is implemented. The ontology produces a relation in between the terms as well as keywords and the user pretend to search. So the metadata based result would be stored inside the Knowledge Database. Inside the stored data, the proposed method introducing a Rank based peer to peer searching strategy; hence the expanded Meta based search results should be prioritized and proper result should be published to the user.

**Peer Rank and Selection Algorithm:**

The peer selection algorithm returns a ranked set of peers. The rank value is equal to the similarity value given by the similarity function. From this set of ranked peers one can, for example, select the best n peers, or all peers whose rank value is above a certain threshold etc a peer selection algorithm [7] is also presented that allows a recommender system peer to select a set of other peers to cooperate with. The algorithm selection mechanism not only ensures a high degree of user satisfaction to the generated recommendation, it also makes sure that every peer behavior has been fairly absorbed. The figure 5 shows the model of the proposed method.

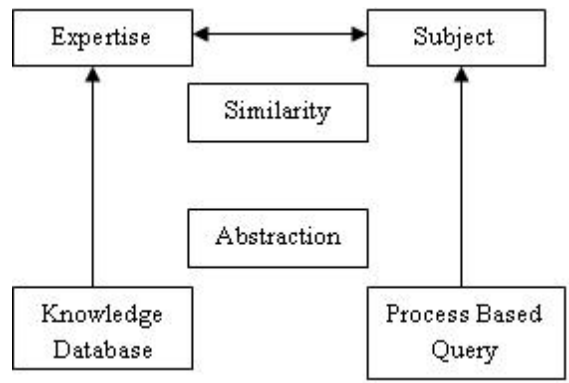

Fig-5 Expertise based selection in the proposed approach

Let P be the probability of spanning forest (the spanning forest will be the total data available inside peer Knowledge Based database)

$$\text{Peer (A)} = \text{Peer (B)} + \text{Peer(C)} + \text{Peer (D)}$$

Here, a small universe of four Peer to Peer networks: **Peer1**, **Peer2**, **Peer3** and **Peer 4** are taken for example. The initial approximation of Page Rank would be evenly divided between these four Peers. Hence, each document would begin with an estimated Page Rank of 0.25.

$$\text{Pp (peer1)} = \text{Pp (peer2)} + \text{Pp (peer3)} + \text{Pp (peer4)}$$

P-Probability of each peer hit selection.
p- Pages that available inside a knowledge database.

In the original form of Page rank algorithm strategies [4], initial values were simply 1 or n-2. This meant that the sum of all Peers was the total number of peers available inside the Knowledge database. Later versions of Page Rank (The formula as mentioned above) would assume a probability distribution between 0.1 and n-1. Here a simple probability distribution will be used along with an initial value 0.5 to 0.2.

As the below mentioned diagram [figure 6] shows priority based selection as well as sorting of data inside a knowledge database [5, 6]. The searched data inside the knowledge database would be prioritized and ordered set of result to be produced for the user.

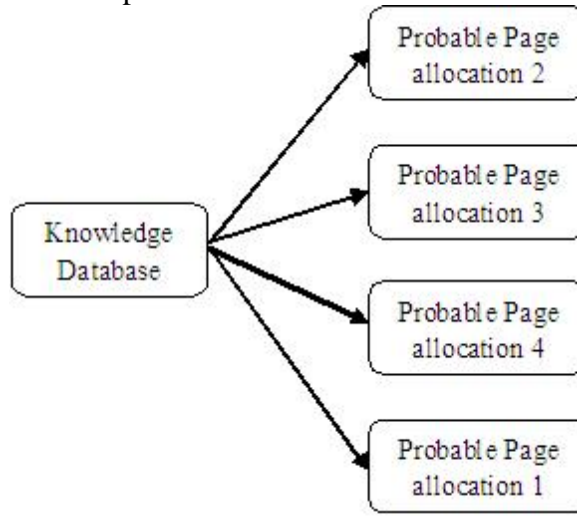

Fig-6 Shows priority selection under the Knowledge Database

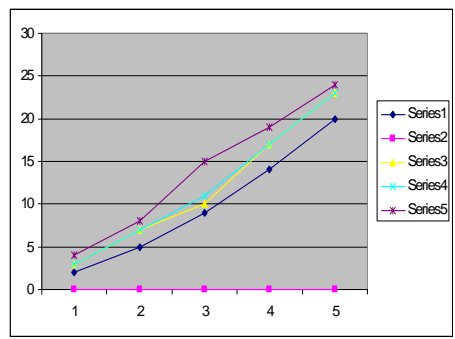

Fig-7 Precision peers

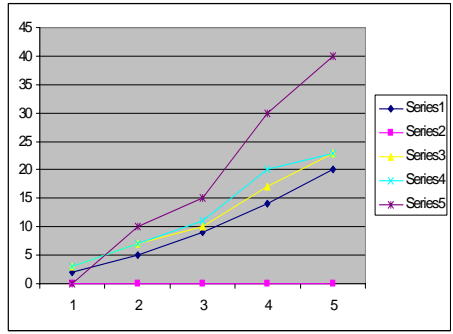

Fig-8 Recall peers

The result of naive distribution will be 0.03% for the topic distribution and 1.3% for the proceedings distribution. The naive approach of random peer selection gives a constant low precision. There will be low recall of peers and documents and there would be also higher number of messages per query. But with the expertise based selection, the precision can be improved considerably. With the expertise based selection the precision of the peer selection (Figure 7) can be improved from 0.03% to 0.15% for the topic distribution and from 1.3% to 15% for the proceedings distribution [5, 6]. The recall number of peers and documents also considerably raised (Figure 8) and the number of messages per query can be reduced. In simulation experiments the selection based with ontology based matching and semantic topologies are carried out and shown how they outperform than mere naive selection models. The result shows that the searching efficiency and Meta data based result is improved compare to the previous model.

**Method-3: (Super peer based Search)**

In unstructured P2P, the flooding method is used as major method for searching the query [11]. Last few methods [8, 9, 10] are based on the flooding with super peers. The main difference between the proposed technique and other super peer methods is: Super peer only perform the function with the help of neighbour super peers. Super peer maintain the two tables:
- ✓ Local table with all information like peers id, route path which obtained last search, relevant information (keywords or Meta data)
- ✓ Global table with all information like neighbour super peer list, route path which obtained last search, relevant information (keywords or Meta data)

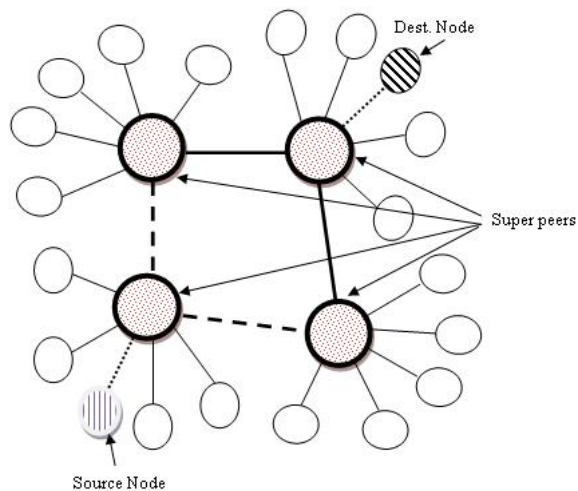

Figure-9. Communication through Super Peers

In the figure 9, the source node sends a query to the super node. Super node receives the query and checks its local table for information by using route path and stored keywords or Meta data. If found the forward the query to the particular node. Else it checks the global table and forwards the query to its neighbour super peers.

The neighbour super peers checks the same as the above method until it reaches the destination node. The TTL value is set only in the super peers and checks for the exhausting. If TTL exhausts then it will repeat the same. The proposed algorithm is:

Step1: Peer node sends a query to super peer.

Step2: Super peer check whether the query is from the same group or not. If it is from same group then, check the cache table in which there will be any peer in the same group having the information or not. If found then, forward the request to the particular peer. Otherwise forward the information to the neighbouring super peers which already cached in the route path table.

Step3: That neighbouring super peer checks the same until it reaches the information by the step 2. If found store the route path and send the information to the source super peer group.

Step4: The TTL value will be checked by the Super peer and resend the query

The average network traffic per query is used to measure the query cost. The average query response time is used to measure the user perceived query quality, i.e., how long a user has to wait before a query result can be sent back to a querying node. The simulation results are shown in Fig. 10 and Fig. 11. The comparison in Fig. 11 shows that, with the help of the proposed algorithm, the average volume of traffic per query decreases. More than 40 percent traffic can be saved by utilizing the super peer. The proposed algorithm can significantly reduce the query cost.

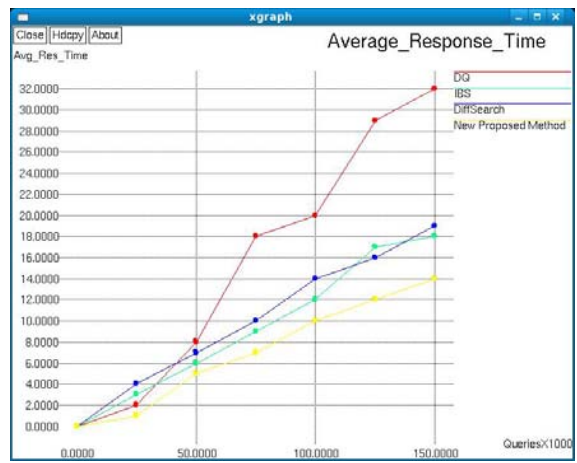

Figure10. Average Response Time for the methods DQ, IBS, DiffSearch and Proposed method

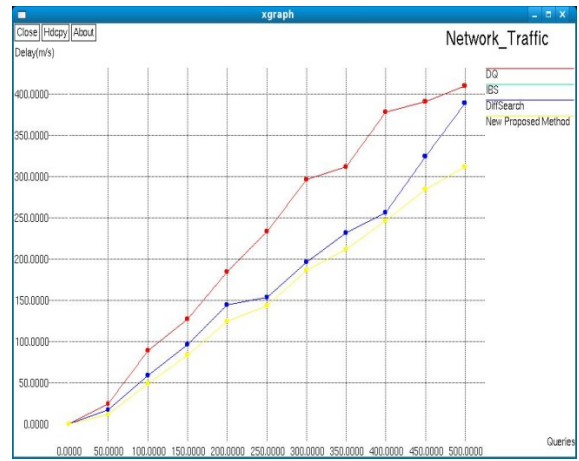

Figure11. Network Traffic for the methods DQ, IBS, DiffSearch and Proposed method

Search methods for decentralized unstructured P2P networks are categorized in different ways. The super peer based blind search is to be improved because of unnecessary traffic can be avoided. The proposed method gives the new approach for the super peer based search with improved efficiency compare to the other methods. The results shown that, the average response time and network traffic can be reduced compare to other methods. Hence, the efficiency of the peer to peer search in the unstructured environment is improved through the new proposed method. The performance of the unstructured environment improves indirectly the traffic also reduced.

**3. Conclusion**

Peer-to-Peer systems are one of the effective ways of communicating and sharing data equally. The new age protocols are able to route efficiently with faster download, better stability and a good user interface. The objective of a search mechanism is to successfully locate resources while incurring low overhead and delay. The performances of the P2P network can be improved by any one of the methods suggested above are proved by the results showed. Hence, the performance of the P2P improves automatically the traffic can be reduced.